\documentstyle[11pt]{article}
\bibliography{plain}
\title{\bf Lewenstein-Sanpera decomposition of a generic $2\times 2$
density matrix by using Wootters's basis } \vspace{20mm}
\author{
  S. J. Akhtarshenas$^{a,b,c}$
\thanks{E-mail:akhtarshenas@tabrizu.ac.ir}
 , M. A. Jafarizadeh$^{a,b,c}$ \thanks{E-mail:jafarizadeh@tabrizu.ac.ir}
\\
\\
$^a${\small Department of Theoretical Physics and Astrophysics,
Tabriz University, Tabriz 51664, Iran.} \\
$^b${\small Institute for Studies in Theoretical Physics and
Mathematics,
 Tehran 19395-1795, Iran.} \\
$^c${\small Research Institute for Fundamental Sciences, Tabriz
51664, Iran.}} \pagebreak

\begin{document}
\maketitle \vspace{15mm}
\newpage
\begin{abstract}
The Lewenstein-Sanpera decomposition for a generic two-qubit
density matrix is obtained  by using Wootters's basis. It is shown
that the average concurrence of the decomposition is equal to the
concurrence of the state.  It is also shown that all the
entanglement content of the state is concentrated in the
Wootters's state $\left|x_1\right>$ associated with the largest
eigenvalue $\lambda_1$ of the Hermitian matrix
$\sqrt{\sqrt{\rho}\tilde{\rho}\sqrt{\rho}}$ . It is shown that a
given density matrix $\rho$ with corresponding set of positive
numbers $\lambda_i$ and Wootters's basis can transforms under
$SO(4,c)$ into a generic $2\times2$ matrix with the same set of
positive numbers but with new Wootters's basis, where the local
unitary transformations correspond to $SO(4,r)$ transformations,
hence, $\rho$ can be represented as coset space $SO(4,c)/SO(4,r)$
together with positive numbers $\lambda_i$. By giving an explicit
parameterization we characterize a generic orbit of group of local
unitary transformations.


{\bf PACs Index: 03.65.Ud }
\end{abstract}
\pagebreak

\vspace{7cm}

\section{Introduction}
Perhaps, quantum entanglement is the most non-classical features
of quantum mechanics \cite{EPR,shcro} which has recently been
attracted much attention although it was discovered many decades
ago by Einstein and  Schr\"{o}dinger \cite{EPR,shcro}. It plays a
central role in quantum information theory and provides potential
resource for quantum communication and information processing
\cite{ben1,ben2,ben3}. Entanglement is usually arise from quantum
correlations between separated subsystems which can not be created
by local actions on each subsystems. By definition, a bipartite
mixed state $\rho$ is said to be separable if it can be expressed
as $$ \rho=\sum_{i}w_{i}\,\rho_i^{(1)}\otimes\rho_i^{(2)},\qquad
w_i\geq 0, \quad \sum_{i}w_i=1, $$ where $\rho_i^{(1)}$ and
$\rho_i^{(2)}$ denote density matrices of subsystems 1 and 2,
respectively. Otherwise the state is entangled.

The central tasks of quantum information theory is to characterize
and quantify entangled states. A first attempt in characterization
of entangled states has been made by Peres and Horodecki family
\cite{peres,horo}. Peres showed that a necessary condition for
separability of a two partite system is that its partial
transposition be positive. Horodeckis have shown that this
condition is sufficient for separability of composite systems only
for dimensions $2\otimes 2$ and $2 \otimes 3$.

There is also an increasing attention in quantifying entanglement,
particularly for mixed states of a bipartite system, and a number
of measures have been proposed \cite{ben3,ved1,ved2,woot}. Among
them the entanglement of formation has more importance, since it
intends to quantify the resources needed to create a given
entangled state.

An interesting description of entanglement is Lewenstein-Sanpera
decomposition \cite{LS}. Lewenstein and Sanpera in \cite{LS}
showed that any two partite density matrix can be represented
optimally as a sum of a separable state and an entangled state.
They have also shown that for  two qubit systems the decomposition
reduces to a mixture of a mixed separable state and an entangled
pure state, thus all non-separability content of the state is
concentrated in the pure entangled state. This leads to an
unambiguous measure of entanglement for any two qubit state as
entanglement of pure state multiplied by the weight of pure part
in the decomposition.

The numerical method for finding the BSA has been reported in Ref.
\cite{LS}. Also in two qubit systems some analytical results for
special states were found in \cite{englert}. An attempt to
generalize the results of Ref. \cite{LS} is made in \cite{karnas}.
In \cite{kus} an algebraic approach to find BSA of a two qubit
state is attempted. They have also shown that the weight of the
entangled part in the decomposition is equal to the concurrence of
the state. In \cite{jaf} we have obtained an analytical expression
for L-S decomposition of Bell decomposable (BD) states. We have
also obtained the optimal decomposition for a class of states
obtained from BD states via some LQCC actions.

Wootters  in \cite{woot} has shown that for any two qubit density
matrix $\rho$ there always exist a decomposition
$\rho=\sum_i\left|x_i\right>\left<x_i\right|$ such that
$\left<x_i|\tilde{x}_j\right>=\lambda_i\delta_{ij}$, where
$\lambda_i$ are square roots of eigenvalues, in decreasing order,
of the non-Hermitian matrix $\rho\tilde{\rho}$. Based on this the
concurrence of the mixed state $\rho$ is defined by
$\max(0,\lambda_1-\lambda_2-\lambda_3-\lambda_4)$ \cite{woot}.

In this paper, by using Wootters's basis, we give an analytical
expression for optimal Lewenstein-Sanpera decomposition for any
two qubit density matrix. We  show that all entanglement content
of the state is concentrated  in the Wootters's state
$\left|x_1\right>$ associated with the largest eigenvalue
$\lambda_1$. It is also shown that  the average concurrence of the
decomposition is equal to the concurrence of the state.

It is shown that a given density matrix $\rho$ with corresponding
set of positive numbers $\lambda_i$ and Wootters's basis can
transforms under $SO(4,c)$ into a generic $2\times2$ matrix with
the same set of positive numbers but with new Wootters's basis,
where the local unitary transformations correspond to $SO(4,r)$
transformations, hence, $\rho$ can be represented as coset space
$SO(4,c)/SO(4,r)$ together with positive numbers $\lambda_i$. By
giving an explicit parameterization we characterize a generic
orbit of group of local unitary transformations.

The paper is organized as follows. In section 2 we review
concurrence for two qubit density matrix. In section 3 we give an
analytical expression for L-S decomposition of a generic density
matrix. We prove that the decomposition is optimal.
Characterization of  the density matrix in terms of orthogonal
group is presented in section 4. Section 4 is devoted to  explicit
parameterization of a generic density matrix up to a local unitary
transformation. The paper is ended with a brief conclusion in
section 5.

\section{Concurrence}
In this section we first review concurrence of two qubit mixed
states. From the various measures proposed to quantify
entanglement, the entanglement of formation has a special position
which in fact intends to quantify the resources needed to create a
given entangled state \cite{ben3}. In the case of pure state if
the density matrix obtained from partial trace over other
subsystems is not pure the state is entangled. For the pure state
$\left|\psi\right>$ of a bipartite system, entropy of the density
matrix associated with either of the the two subsystems is a good
measure of entanglement
$$
E(\psi)=-Tr(\rho_A\log_2\rho_A)=-Tr(\rho_B\log_2\rho_B),
$$
where $\rho_A=Tr_B(\left|\psi\right>\left<\psi\right|)$ and
$\rho_B$ defined similarly. Due to classical correlations where
exist in the mixed state each subsystem can have non-zero entropy
even if there is no entanglement, therefore von Neumann entropy of
a subsystem is no longer a good measure of entanglement. For a
mixed state entanglement of formation is defined as the minimum of
average entropy of the state over all pure state decompositions of
the state
\begin{equation}\label{EoF}
E_f(\rho)=\min\sum_i p_i E(\psi_i).
\end{equation}

Wootters in \cite{woot} has shown that for a two qubit system
entanglement of formation of a mixed state $\rho$ can be defined
as
\begin{equation}
E_f(\rho)=H\left(\frac{1}{2}+\frac{1}{2}\sqrt{1-C^2}\right),
\end{equation}
where $H(x)=-x\ln{x}-(1-x)\ln{(1-x)}$ is binary entropy and
concurrence $C(\rho)$ is defined by
\begin{equation}\label{concurrence}
C(\rho)=\max\{0,\lambda_1-\lambda_2-\lambda_3-\lambda_4\},
\end{equation}
where the $\lambda_i$ are the non-negative eigenvalues, in
decreasing order, of the Hermitian matrix
$R\equiv\sqrt{\sqrt{\rho}{\tilde \rho}\sqrt{\rho}}$ and
\begin{equation}\label{rhotilde}
{\tilde \rho}
=(\sigma_y\otimes\sigma_y)\rho^{\ast}(\sigma_y\otimes\sigma_y),
\end{equation}
where $\rho^{\ast}$ is the complex conjugate of $\rho$ when it is
expressed in a standard basis such as
$\{\left|\uparrow\uparrow\right>,
\left|\uparrow\downarrow\right>\},\{\left|\downarrow\uparrow\right>,
\left|\downarrow\downarrow\right>\}$ and $\sigma_y$ represent
Pauli matrix in local basis $\{\left|\uparrow\right>,
\left|\downarrow\right>\}$ .

Consider a generic two qubit density matrix $\rho$ with its
subnormalized orthogonal eigenvectors $\left|v_i\right>$, i.e.
$\rho=\sum_i \left|v_i\right>\left<v_i\right|$. There always exist
a decomposition \cite{woot}
\begin{equation}\label{rhox}
\rho=\sum_i\left|x_i\right>\left<x_i\right|
\end{equation}
where Wootters states $\left|x_i\right>$ are defined by
\begin{equation}\label{xvector}
\left|x_i\right>=\sum_{j}^{4}U_{ij}^{\ast}\left|v_i\right>, \qquad
\mbox{for}\quad i=1,2,3,4,
\end{equation}
such that
\begin{equation}\label{xortho}
\left<x_i\mid \tilde{x}_j\right>=(U\tau
U^T)_{ij}=\lambda_i\delta_{ij},
\end{equation}
where $\tau_{ij}=\left<v_i\mid \tilde{v}_j\right>$ is a symmetric
but not necessarily Hermitian matrix. To construct
$\left|x_i\right>$ we use the fact that for any symmetric matrix
$\tau$ one can always find a unitary matrix $U$ in such a way that
$\lambda_i$ are real and non-negative, that is, they are the
square roots of eigenvalues of $\tau\tau^{\ast}$ which are the
same as eigenvalues of $R$. Alternatively one can always find $U$
such that $\lambda_i$ appear in decreasing order.

\section{Lewenstein-Sanpera decomposition}
According to Lewenstein-Sanpera decomposition \cite{LS}, any two
qubit  density matrix $\rho$ can be written as
\begin{equation}\label{LSD}
\rho=\lambda\rho_{sep}+(1-\lambda)\left|\psi\right>\left<\psi\right|,
\quad\quad \lambda\in[0,1],
\end{equation}
where $\rho_{sep}$ is a separable density matrix and
$\left|\psi\right>$ is a pure entangled state. The
Lewenstein-Sanpera decomposition of a given density matrix $\rho$
is not unique and, in general, there is a continuum set of L-S
decomposition to choose from. The optimal decomposition is,
however, unique for which $\lambda$ is maximal, and
\begin{equation}\label{LSDopt}
\rho=\lambda^{(opt)}\rho_{sep}^{(opt)}
+(1-\lambda^{(opt)})|\psi^{(opt)}\left>\right<\psi^{(opt)}|\;,
\quad\quad \lambda^{(opt)}\in[0,1].
\end{equation}
 In Ref. \cite{LS}    Lewenstein and Sanpera have shown that any other
  decomposition of the form
 $\rho={\tilde \lambda}{\tilde \rho}_{sep}
 +(1-{\tilde \lambda})|{\tilde \psi}\left>\right<{\tilde \psi}|$
 , with ${\tilde \lambda}\in[0,1]$ such that ${\tilde
 \rho}\neq\rho^{(opt)}$ necessarily implies that ${\tilde
 \lambda}<\lambda^{(opt)}$ \cite{LS}.

Here in this section we obtain L-S decomposition for a generic two
qubit density matrix by using Wootters states. First we define
states $\left|x^\prime_i \right>$ in terms of Wootters states as
\begin{equation}\label{xprime}
\left|x^\prime_i \right>=\frac{\left|x_i
\right>}{\sqrt{\lambda_i}},\qquad \mbox{for}\,i=1,2,3,4.
\end{equation}
Thus the decomposition given in Eq. (\ref{rhox}) becomes
\begin{equation}\label{rhoxp}
\rho=\sum_i\lambda_i\left|x^{\prime}_i\right>\left<x^{\prime}_i\right|,
\end{equation}
where can be rewritten in the following form
\begin{equation}\label{LSD1}
\begin{array}{rl}
\rho & =\sum_{i=1}^{4}\lambda_i\left|x^\prime_i\right>\left<x^\prime_i\right| \\
& =(\lambda_1-\lambda_2-\lambda_3-\lambda_4)
\left|x^\prime_1\right>\left<x^\prime_1\right|
+(\lambda_2+\lambda_3+\lambda_4)\left|x^\prime_1\right>\left<x^\prime_1\right|
+\sum_{j=2}^4\lambda_j\left|x^\prime_j\right>\left<x^\prime_j\right|
\\
&
=(1-\lambda)\left|\psi\right>\left<\psi\right|+\lambda\,\rho_{sep}\,,
\end{array}
\end{equation}
where separable density matrix $\rho_{sep}$ and entangled pure
state $\left|\psi\right>$ are given by
\begin{equation}\label{rhosep}
\rho_{sep}=\left(\frac{\lambda_2+\lambda_3+\lambda_4}{\lambda\lambda_1}\right)
\left|x_1\right>\left<x_1\right|
+\frac{1}{\lambda}\sum_{j=2}^4\left|x_j\right>\left<x_j\right|,
\end{equation}
and
\begin{equation}\label{psi}
\left|\psi\right>=\frac{\left|x_1\right>}{\sqrt{\left<x_1|x_1\right>}},
\end{equation}
respectively, and parameter $\lambda$ is equal to
\begin{equation}\label{lambda}
\lambda=1-\left(\frac{\lambda_1-\lambda_2-\lambda_3-\lambda_4}{\lambda_1}\right)
\left<x_1|x_1\right>.
\end{equation}
Equation (\ref{psi}) shows that all entanglement content of the
state is concentrated in the Wootters state $\left|x_1\right>$
associated  to the largest eigenvalues $\lambda_1$. It is also
worth to note that average concurrence of the decomposition is
\begin{equation}\label{avecon}
(1-\lambda)\left<\psi|\tilde{\psi}\right>
=(\lambda_1-\lambda_2-\lambda_3-\lambda_4),
\end{equation}
that is,  it is equal to the concurrence of the state.

In order to show that the decomposition given in (\ref{LSD1}) is
optimal we first show that  $\rho_{sep}$ can be represented as a
convex sum of product states. First note that $\rho_{sep}$  can be
written as
\begin{equation}\label{rhosepxpp}
\rho_{sep}=\sum_{i=1}^{4}\left|x^{\prime\prime}_i\right>\left<x^{\prime\prime}_i\right|,
\end{equation}
where $\left|x^{\prime\prime}_i\right>$ are defined by
\begin{equation}\label{xpp}
\left|x^{\prime\prime}_1\right>=\sqrt{\frac{\lambda_2+\lambda_3+\lambda_4}{\lambda\,\lambda_1}}
\left|x_1\right>,
\qquad\left|x^{\prime\prime}_j\right>=\frac{1}{\sqrt{\lambda}}\left|x_j\right>,\quad
\mbox{for}\,j=2,3,4.
\end{equation}
 Obviously, the basis
$\left|x^{\prime\prime}_j\right> (i=1,2,3,4)$ satisfy the
following relations

$$\left<x^{\prime\prime}_i|\tilde{x^{\prime\prime}_j}\right>
=\lambda^{\prime\prime}_i\delta_{ij},$$ where
$\left|\tilde{x^{\prime\prime}_j}\right> (j=1,2,3,4)$ are
corresponding dual basis.

 Wootters in \cite{woot} has shown that
any two qubit separable density matrix with decomposition given in
Eq. (\ref{rhosepxpp}) can be expanded in terms of following
product states
\begin{equation}\label{z1234-1}
\left|z_1\right>=\frac{1}{2}\left(e^{i\theta_1}\left|x^{\prime\prime}_1\right>
+e^{i\theta_2}\left|x^{\prime\prime}_2\right>+e^{i\theta_3}\left|x^{\prime\prime}_3\right>
+e^{i\theta_4}\left|x^{\prime\prime}_4\right>\right),
\end{equation}
\begin{equation}\label{z1234-1}
\left|z_2\right>=\frac{1}{2}\left(e^{i\theta_1}\left|x^{\prime\prime}_1\right>
+e^{i\theta_2}\left|x^{\prime\prime}_2\right>-e^{i\theta_3}\left|x^{\prime\prime}_3\right>
-e^{i\theta_4}\left|x^{\prime\prime}_4\right>\right),
\end{equation}
\begin{equation}\label{z1234-1}
\left|z_3\right>=\frac{1}{2}\left(e^{i\theta_1}\left|x^{\prime\prime}_1\right>
-e^{i\theta_2}\left|x^{\prime\prime}_2\right>+e^{i\theta_3}\left|x^{\prime\prime}_3\right>
-e^{i\theta_4}\left|x^{\prime\prime}_4\right>\right),
\end{equation}
\begin{equation}\label{z1234-1}
\left|z_4\right>=\frac{1}{2}\left(e^{i\theta_1}\left|x^{\prime\prime}_1\right>
-e^{i\theta_2}\left|x^{\prime\prime}_2\right>-e^{i\theta_3}\left|x^{\prime\prime}_3\right>
+e^{i\theta_4}\left|x^{\prime\prime}_4\right>\right),
\end{equation}
where zero concurrence is guaranteed with
$\sum_{j=1}e^{2i\theta_j}\lambda^{\prime\prime}_j=0$.

Now using the fact that the eigenvalues $\lambda^{\prime\prime}_i$
of boundary separable states $\rho_{sep}$ satisfy constraint
$\lambda^{\prime\prime}_1-\lambda^{\prime\prime}_2
-\lambda^{\prime\prime}_3-\lambda^{\prime\prime}_4=0$, we can
choose phase factors $\theta_i$ as
$\theta_2=\theta_3=\theta_4=\theta_1+\pi$. Choosing $\theta_1=0$
we arrive at the following product ensemble for $\rho_{sep}$
\begin{equation}\label{z1-2}
\left|z_1\right>=\frac{1}{2\sqrt{\lambda}}\left(
\sqrt{\frac{\lambda_2+\lambda_3+\lambda_4}{\lambda_1}}\left|x_1\right>
-i\left|x_2\right>-i\left|x_3\right>-i\left|x_4\right>\right),
\end{equation}
\begin{equation}\label{z2-2}
\left|z_2\right>=\frac{1}{2\sqrt{\lambda}}\left(
\sqrt{\frac{\lambda_2+\lambda_3+\lambda_4}{\lambda_1}}\left|x_1\right>
-i\left|x_2\right>+i\left|x_3\right>+i\left|x_4\right>\right),
\end{equation}
\begin{equation}\label{z3-2}
\left|z_3\right>=\frac{1}{2\sqrt{\lambda}}\left(
\sqrt{\frac{\lambda_2+\lambda_3+\lambda_4}{\lambda_1}}\left|x_1\right>
+i\left|x_2\right>-i\left|x_3\right>+i\left|x_4\right>\right),
\end{equation}
\begin{equation}\label{z4-2}
\left|z_4\right>=\frac{1}{2\sqrt{\lambda}}\left(
\sqrt{\frac{\lambda_2+\lambda_3+\lambda_4}{\lambda_1}}\left|x_1\right>
+i\left|x_2\right>+i\left|x_3\right>-i\left|x_4\right>\right).
\end{equation}
It can be easily seen that all $\left|z_i\right>$ have zero
concurrence and also $\rho_{sep}$ can be expanded as
\begin{equation}\label{rhosepz}
\rho_{sep}=\sum_{i=1}^{4}\left|z_i\right>\left<z_i\right|.
\end{equation}

In the rest of this section we will prove that the decomposition
(\ref{LSD1}) is the optimal one. To do so we have to find a
decomposition for $\rho_{sep}$ in terms of product states
$\left|e_\alpha,f_\alpha\right>$, i.e.
\begin{equation}
\rho_{sep}=\sum_{\alpha}\Lambda_\alpha
\left|e_\alpha,f_\alpha\right>\left<e_\alpha,f_\alpha\right|
\end{equation}
such that the following conditions are satisfied \cite{LS}

i) All $\Lambda_\alpha$ are maximal with respect to
$\rho_\alpha=\Lambda_\alpha
\left|e_\alpha,f_\alpha\right>\left<e_\alpha,f_\alpha\right|
+(1-\lambda)\left|\psi\right>\left<\psi\right|$ and projector
$P_\alpha=\left|e_\alpha,f_\alpha\right>\left<e_\alpha,f_\alpha\right|$.

ii) All pairs $(\Lambda_\alpha,\Lambda_\beta)$ are maximal with
respect to $\rho_{\alpha\beta}=\Lambda_\alpha
\left|e_\alpha,f_\alpha\right>\left<e_\alpha,f_\alpha\right|
+\Lambda_\beta
\left|e_\beta,f_\beta\right>\left<e_\beta,f_\beta\right|
+(1-\lambda)\left|\psi\right>\left<\psi\right|$ and the pairs of
projector $(P_\alpha,P_\beta)$.

Then according to \cite{LS} $\rho_{sep}$ is BSA and the
decomposition given in Eq. (\ref{LSD1}) is optimal.

Lewenstein and Sanpera in \cite{LS} have shown that
$\Lambda_{\alpha}$ is maximal with respect to $\rho_{\alpha}$ and
$P_{\alpha}=\left|e_\alpha,f_\alpha\right>\left<\psi\right|$ iff

a) if $\left|e_\alpha,f_\alpha\right>\not\in {\cal
R}(\rho_{\alpha})$ then $\Lambda_{\alpha}=0$, and b) if
$\left|e_\alpha,f_\alpha\right>\in {\cal R}(\rho_\alpha)$ then
$\Lambda_{\alpha}=
\left<e_\alpha,f_\alpha\right|\rho_{\alpha}^{-1}\left|e_\alpha,f_\alpha\right>^{-1}>0$.
They have also shown that a pair $(\Lambda_1,\Lambda_2)$ is
maximal with respect to $\rho_{12}$ and a pair of projectors
$(P_1,P_2)$ iff:

a) if $\left|e_1,f_1\right>$, $\left|e_2,f_2\right>$ do not belong
to ${\cal R}(\rho_{12})$ then $\Lambda_1=\Lambda_2=0$; b) if
$\left|e_1,f_1\right>$ does not belong, while
$\left|e_2,f_2\right>\in{\cal R}(\rho_{12})$ then $\Lambda_1=0$,
$\Lambda_2=\left<e_2,f_2\right|\rho_{12}^{-1}\left|e_2,f_2\right>^{-1}$;
c) if $\left|e_1,f_1\right>$, $\left|e_2,f_2\right>\in {\cal
R}(\rho_{12})$ and
$\left<e_1,f_1\right|\rho_{12}^{-1}\left|e_2,f_2\right>=0$ then
$\Lambda_i=\left<e_i,f_i\right|\rho_{12}^{-1}\left|e_i,f_i\right>^{-1}$,
$i=1,2$; d) finally, if $\left|e_1,f_1\right>,
\left|e_2,f_2\right>\in {\cal R}(\rho_{12})$ and
$\left<e_1,f_1\right|\rho_{12}^{-1}\left|e_2,f_2\right>\neq 0$
then

\begin{equation}\label{Lambda12}
\begin{array}{lr}
\Lambda_1=
&(\left<e_2,f_2\right|\rho_{12}^{-1}\left|e_2,f_2\right>
-\mid\left<e_1,f_1\right|\rho_{12}^{-1}\left|e_2,f_2\right>\mid)/D, \\
\Lambda_2=
&(\left<e_1,f_1\right|\rho_{12}^{-1}\left|e_1,f_1\right>
-\mid\left<e_1,f_1\right|\rho_{12}^{-1}\left|e_2,f_2\right>\mid)/D,
\end{array}
\end{equation}
where
$D=\left<e_1,f_1\right|\rho_{12}^{-1}\left|e_1,f_1\right>\left<e_2,f_2\right|
\rho_{12}^{-1}\left|e_2,f_2\right>
-\mid\left<e_1,f_1\right|\rho_{-1}\left|e_2,f_2\right>\mid^2$.

Now we return to show that the decomposition given in (\ref{LSD1})
is optimal. We first consider the cases that $\rho$ has full rank.
Let us consider the set of four product vectors
$\{\left|z_i\right>\}$ and one entangled state $\left|x_1\right>$.
In Ref. \cite{woot} it is shown that the ensemble
$\{\left|z_i\right>\}$ are linearly independent, also it is
straightforward to see that three vectors $\left|z_\alpha\right>$
, $\left|z_\beta\right>$ and $\left|x_1\right>$ are linearly
independent. Now let us consider matrices
$\rho_\alpha=\Lambda_\alpha \left|
z_\alpha\right>\left<z_\alpha\right|
+(1-\lambda)\left|x_1\right>\left<x_1\right|$. Due to independency
of $\left|z_\alpha\right>$ and $\left|x_1\right>$ we can deduce
that the range of $\rho_\alpha$ is two dimensional, thus after
restriction to its range and defining dual basis $\left|{\hat
z}_\alpha\right>$ and $\left|{\hat x}_1\right>$ we can expand
restricted inverse $\rho_\alpha^{-1}$ as
$\rho_{\alpha}^{-1}=\Lambda_\alpha^{-1}|{\hat
z}_\alpha\left>\right<{\hat z}_\alpha| +(1-\lambda)^{-1}|{\hat
x}_1\left>\right<{\hat x}_1|$ (see appendix). Using Eq. (\ref{A1})
it is easy to see that $\left<z_\alpha|\rho_{\alpha}^{-1}
|z_\alpha\right>=\Lambda_\alpha^{-1}$. This shows that
$\Lambda_\alpha$ are maximal with respect to $\rho_\alpha$ and the
projector $P_\alpha=\left|z_\alpha\right>\left<z_\alpha\right|$.

Similarly considering matrices $\rho_{\alpha\beta}=\Lambda_\alpha
\left|z_\alpha\right>\left<z_\alpha\right| +\Lambda_\beta
\left|z_\beta\right>\left<z_\beta\right|
+(1-\lambda)\left|x_1\right>\left<x_1\right|$ and considering the
independency of  vectors $\left|z_\alpha\right>$,
$\left|z_\beta\right>$ and $\left|x_1\right>$ we see that the rang
of $\rho_{\alpha\beta}$ is three dimensional where after
restriction to its range and defining their dual basis
$\left|{\hat z}_\alpha\right>$, $\left|{\hat z}_\beta\right>$ and
$\left|{\hat x}_1\right>$ we can write restricted inverse
$\rho_{\alpha\beta}^{-1}$ as
$\rho_{\alpha\beta}^{-1}=\Lambda_\alpha^{-1}|{\hat
z}_\alpha\left>\right<{\hat z}_\alpha|+\Lambda_\beta^{-1}|{\hat
z}_\beta\left>\right<{\hat z}_\beta|+(1-\lambda)^{-1}|{\hat
x}_1\left>\right<{\hat x}_1|$. Then it is straightforward to get
$\left<z_\alpha\right|\rho_{\alpha\beta}^{-1}\left|
z_\alpha\right>=\Lambda_\alpha^{-1}$,
$\left<z_\beta\right|\rho_{\alpha\beta}^{-1}\left|
z_\beta\right>=\Lambda_\beta^{-1}$ and $\left<
z_\alpha\right|\rho_{\alpha\beta}^{-1}\left|z_\beta\right>=0$.
This implies that the pair $(\Lambda_\alpha,\Lambda_\beta)$ are
maximal with respect to $\rho_{\alpha\beta}$ and the pair of
projectors $(P_\alpha,P_\beta)$, thus complete the proof that the
decomposition given in Eq. (\ref{LSD1}) is optimal.

We now consider the cases that $\rho$ has rank three, that is
$\lambda_4=0$. In this case the pairs
$\{\left|z_1\right>,\left|z_4\right>\}$ and also
$\{\left|z_2\right>,\left|z_3\right>\}$ are no longer independent
with respect to $\left|x_1\right>$. In former case we can evaluate
$\left|x_1\right>$ in terms of $\left|z_1\right>$ and
$\left|z_4\right>$ then matrix $\rho_{14}$ can be written in terms
of two basis $\left|z_1\right>$ and $\left|z_4\right>$ which
yields after some calculations,
$\left<z_1|\rho_{14}^{-1}|z_1\right>= \frac{1}{\Gamma_{14}}
\left(\Lambda_4+(1-\lambda)\left(\frac{\lambda_1}{\lambda_2+\lambda_3}\right)\right)$,
$\left<z_4|\rho_{14}^{-1}|z_4\right>= \frac{1}{\Gamma_{14}}
\left(\Lambda_1+(1-\lambda)\left(\frac{\lambda_1}{\lambda_2+\lambda_3}\right)\right)$
 and $\left<z_1|\rho_{14}^{-1}|z_4\right>=
\frac{-1}{\Gamma_{14}}
\left((1-\lambda)\left(\frac{\lambda_1}{\lambda_2+\lambda_3}\right)\right)$
where $\Gamma_{14}=\left(\Lambda_1\Lambda_4
+(\Lambda_1+\Lambda_2)(1-\lambda)\left(\frac{\lambda_1}{\lambda_2+\lambda_3}\right)\right)
$. Using the above results together with Eqs. (\ref{Lambda12}) we
obtain the maximality of pair $(\Lambda_1,\Lambda_{4})$ with
respect to $\rho_{14}$ and the pair of projectors $(P_1,P_{4})$.
Similarly in the second case one can express $\left|x_1\right>$ in
terms of $\left|z_2\right>$ and $\left|z_3\right>$ and evaluate
$\rho_{23}^{-1}$, which get $\left<z_2|\rho_{23}^{-1}|z_2\right>=
\frac{1}{\Gamma_{23}}
\left(\Lambda_3+(1-\lambda)\left(\frac{\lambda_1}{\lambda_2+\lambda_3}\right)\right)$,
$\left<z_3|\rho_{23}^{-1}|z_3\right>= \frac{1}{\Gamma_{23}}
\left(\Lambda_2+(1-\lambda)\left(\frac{\lambda_1}{\lambda_2+\lambda_3}\right)\right)
$ and $\left<z_2|\rho_{23}^{-1}|z_3\right>=\frac{-1}{\Gamma_{23}}
\left((1-\lambda)\left(\frac{\lambda_1}{\lambda_2+\lambda_3}\right)\right)$
with $\Gamma_{23}=\left(\Lambda_2\Lambda_3
+(\Lambda_2+\Lambda_3)(1-\lambda)\left(\frac{\lambda_1}{\lambda_2+\lambda_3}\right)\right)
$ , together with Eqs. (\ref{Lambda12}) we obtain the maximality
of pair $(\Lambda_2,\Lambda_{3})$ with respect to $\rho_{23}$ and
the pair of projectors $(P_2,P_{3})$. For other choices of
$\alpha$ and $\beta$ three vectors $\left|z_\alpha\right>$
,$\left|z_\beta\right>$ and $\left|x_1\right>$ remain linearly
independent thus we can prove maximality of pairs
$(\Lambda_\alpha,\Lambda_\beta)$ in the same way that we proved it
in full rank case.

Finally let us consider cases that $\rho$ has rank two, that is
$\lambda_3=\lambda_4=0$. In this case we have
$\left|z_1\right>=\left|z_2\right>$ and
$\left|z_3\right>=\left|z_4\right>$. It is now sufficient to take
$\left|z_1\right>$ and $\left|z_3\right>$ as product ensemble. But
in this case vectors $\left|z_1\right>$ and $\left|z_3\right>$ are
not independent with respect to $\left|x_1\right>$. We express
$\left|x_1\right>$ in terms of $\left|z_1\right>$ and
$\left|z_3\right>$ then matrix $\rho_{13}$ can be written in terms
of two vectors $\left|z_1\right>$ and $\left|z_3\right>$ and after
some calculations we get $\left<z_1|\rho_{13}^{-1}|z_1\right>=
\frac{1}{\Gamma_{13}}
\left(\Lambda_3+(1-\lambda)\left(\frac{\lambda_1}{\lambda_2}\right)\right)
$, $\left<z_3|\rho_{13}^{-1}|z_3\right>=\frac{1}{\Gamma_{13}}
\left(\Lambda_1+(1-\lambda)\left(\frac{\lambda_1}{\lambda_2}\right)\right)
$ and $\left<z_1|\rho_{13}^{-1}|z_3\right>= \frac{-1}{\Gamma_{13}}
\left((1-\lambda)\left(\frac{\lambda_1}{\lambda_2}\right)\right)$
where $\Gamma_{13}= \left(\Lambda_1\Lambda_3+
(\Lambda_1+\Lambda_3)(1-\lambda)\left(\frac{\lambda_1}{\lambda_2}\right)\right)$.
Using the above results together with Eqs. (\ref{Lambda12}) we
obtain the maximality of pairs $(\Lambda_1,\Lambda_3)$ with
respect to $\rho_{13}$ and the pairs of projectors $(P_1,P_3)$.

\section{Coset structure for a generic $2 \times 2 $ density matrix in Wootters's basis}
In this section we obtain an explicit parameterization for a
generic two qubit density matrix in Wootters's basis. To this aim
for any density matrix $\rho$ with decomposition given in Eq.
(\ref{rhoxp}) we define matrix $X$ as
\begin{equation}\label{X}
X=\left(\left|x^\prime_1\right>, \left|x^\prime_2\right>,
\left|x^\prime_3\right>, \left|x^\prime_4\right> \right).
\end{equation}
Analogously by defining matrix
\begin{equation}\label{Xtilde}
\widetilde{X}=\left(\left|\widetilde{x^\prime}_1\right>,
\left|\widetilde{x^\prime}_2\right>,
\left|\widetilde{x^\prime}_3\right>,
\left|\widetilde{x^\prime}_4\right> \right),
\end{equation}
Eq. (\ref{xortho}) takes the following form
\begin{equation}\label{XtX}
\widetilde{X}^\dag X=X^T\sigma_y\otimes\sigma_y X=I.
\end{equation}
Since matrix  $\sigma_y\otimes\sigma_y$ is symmetric it  can be
diagonalized  as
\begin{equation}\label{sigmayy}
\sigma_y\otimes\sigma_y= O^T \eta^2O,
\end{equation}
where $O$ is an orthogonal matrix defined by
\begin{equation}\label{O}
O=\frac{1}{\sqrt{2}}\left(\begin{array}{cccc}
1 & 0 & 0 & 1 \\
0 & 1 & 1 & 0 \\
0 & 1 & -1 & 0 \\
1 & 0 & 0 & -1
\end{array}\right),
\end{equation}
and $\eta$ is the diagonal matrix
\begin{equation}\label{eta}
\eta=\left(\begin{array}{cccc}
i & 0 & 0 & 0 \\
0 & 1 & 0 & 0 \\
0 & 0 & i & 0 \\
0 & 0 & 0 & 1
\end{array}\right).
\end{equation}

Using Eq. (\ref{sigmayy}) we can rewrite Eq. (\ref{XtX}) as
\begin{equation}\label{YtY}
Y^T Y=I,
\end{equation}
where

\begin{equation}\label{YX}
Y=\eta\, O X.
\end{equation}
Equation (\ref{YtY}) shows that Y is a complex 4-dimensional
orthogonal matrix. This means that a given density matrix $\rho$
with corresponding set of positive numbers $\lambda_i$ and
Wootters's basis can transforms under $SO(4,c)$ into a generic
$2\times2$ density matrix with the same set of positive numbers
but with new Wootters's basis. This implies that the space of two
qubit density matrices can be characterize with 12-dimensional (as
real manifold) space of complex orthogonal group $SO(4,c)$
together with four positive numbers $\lambda_i$. Of course the
normalization condition reduces number of parameters to 15.

As far as entanglement is concerned the states $\rho$ and
$\rho^\prime$ are equivalent if they are on the same orbit of the
group of local transformation, that is, if there exist local
unitary transformation $U_1\otimes U_2$ such that
$\rho^\prime=(U_1\otimes U_2)\rho(U_1\otimes U_2)^\dag$, where
$U_1$ and $U_2$ are unitary transformations acting on Hilbert
spaces of particles $A$ and $B$, respectively.

It can be easily seen that under above mentioned  local unitary
transformations of density matrix $\rho$, the matrix $X$
transforms as
\begin{equation}\label{XprimeX}
X\rightarrow X^\prime=(U_1\otimes U_2)X.
\end{equation}
It is worth to mention that $X^\prime$ also satisfy Eq.
(\ref{XtX}). To show that this is indeed the case, we need to note
that $X^{\prime^T}\sigma_y\otimes\sigma_y X^\prime
=X^T(U_1^T\sigma_y U_1)\otimes(U_2^T\sigma_y U_2)X$. By using
$(\sigma_y)_{ij}=-i\epsilon_{ij}$ we get $(U^T\sigma_y U)_{ij} =
-i\epsilon_{kl}U_{ki}U_{lj}=-i\det(U)\epsilon_{ij}=\sigma_y$,
where the fact that $U_i\in SU(2)$, thus having unit determinant,
have been used. This implies that
\begin{equation}\label{XptXp}
X^{\prime^T}\sigma_y\otimes\sigma_y X^\prime=I.
\end{equation}
By defining $Y^\prime$ as
\begin{equation}\label{YprimeXprime}
Y^\prime=\eta\,OX^\prime,
\end{equation}
one can easily show that $Y^\prime$ is also satisfied
orthogonality condition
\begin{equation}\label{YptYp}
Y^{\prime^T}Y^\prime=I.
\end{equation}
Now by using Eq. (\ref{YprimeXprime}) and inverting Eq.
(\ref{XprimeX}), we can express $Y^\prime$ in terms of $Y$
\begin{equation}\label{YprimeY}
Y^\prime= (\eta\,O)(U_1\otimes U_2)(\eta\,O)^{-1} Y.
\end{equation}
Now by using the fact that $(\eta\,O)\exp({\cal U}_1\otimes {\cal
U}_2)(\eta\,O)^{-1}=\exp((\eta\,O)({\cal U}_1\otimes {\cal
U}_2)(\eta\,O)^{-1})$ and using the explicit form for generators
$({\cal U}_1\otimes {\cal U}_2)$ of local group, one can after
some algebraic calculations see that $(\eta\,O)({\cal U}_1\otimes
{\cal U}_2)(\eta\,O)^{-1}$ is real antisymmetric matrix. This
means that under local unitary transformations matrix $Y$
transforms with $SO(4,r)$ group. So we can parameterize the space
of two qubit density matrices as 6-dimensional coset space
$SO(4,c)/SO(4,r)$ together with 4 positive numbers $\lambda_i$,
which again normailzation condition reduces the number of
parameters to 9.

In the following we will obtain an explicit parameterization for a
generic two qubit density matrix. First note that we can decompose
coset $SO(4,c)/SO(4,r)$ as \cite{gilmore}
\begin{equation}\label{cosetdecom}
\frac{SO(4,c)}{SO(4,r)}=\frac{SO(4,c)/SO(4,r)}{SO(2,c)/SO(2,r)\otimes
SO(2,c)/SO(2,r)}\bigotimes
\left(\frac{SO(2,c)}{SO(2,r)}\otimes\frac{SO(2,c)}{SO(2,r)}\right),
\end{equation}
that is, coset representative $Y$ can be decomposed as $Y=Y_1Y_2$.
One can easily show  that coset representative of
$SO(2,c)/SO(2,r)$ has the following form
\begin{equation}
\exp\left(\begin{array}{cc}
0 & i\phi \\
-i\phi & 0 \end{array} \right)= \left(\begin{array}{cc}
\cosh{\phi} & i\sinh{\phi} \\
-i\sinh{\phi} & \cosh{\phi} \end{array} \right).
\end{equation}
Thus $Y_2$ can be written as
\begin{equation}\label{Y2}
Y_2=\left(\begin{array}{c|c}
\begin{array}{cc}
\cosh{\phi_1} & i\sinh{\phi_1} \\
-i\sinh{\phi_1} & \cosh{\phi_1} \end{array} & 0 \\
\hline 0 & \begin{array}{cc}
\cosh{\phi_2} & i\sinh{\phi_2} \\
-i\sinh{\phi_2} & \cosh{\phi_2} \end{array}
\end{array}\right).
\end{equation}
On the other hand $Y_1$ can be evaluated as
\begin{equation}\label{Y1-1}
Y_1=\exp \left( \begin{array}{c|c} 0 & iB \\ \hline -iB^T & 0
\end{array}\right)=
\left( \begin{array}{c|c} \cosh{\sqrt{BB^T}} & iB\frac{\sinh{\sqrt{B^TB}}}{\sqrt{B^TB}} \\
\hline -i\frac{\sinh{\sqrt{B^TB}}}{\sqrt{B^TB}}B^T
 & \cosh{\sqrt{B^TB}}
\end{array}\right)=\left( \begin{array}{c|c} \sqrt{I+CC^T} & iC \\
\hline
 -iC^T & \sqrt{I+C^TC}
\end{array}\right),
\end{equation}
where $B$ is a $2\times2$ matrix and in the last step we used
$C=B\frac{\sinh{\sqrt{B^TB}}}{\sqrt{B^TB}}$. Now using the
singular value decomposition $C=O_1DO^T_2$, Eq. (\ref{Y1-1})
becomes
\begin{equation}\label{Y1-2}
Y_1=\left( \begin{array}{c|c} O_1\sqrt{I+D^2}O^T_1 & iO_1DO^T_2 \\
\hline
 -iO_2DO^T_1 & O_2\sqrt{I+D^2}O^T_2
\end{array}\right),
\end{equation}
where $D$ is a non-negative diagonal matrix. It can be easily seen
that Eq. (\ref{Y1-2}) can be decomposed as
\begin{equation}\label{Y1-3}
Y_1=\left( \begin{array}{c|c} O_1 & 0 \\
\hline
 0 & O_2
\end{array}\right)
\left( \begin{array}{c|c} \sqrt{I+D^2} & iD \\
\hline
 -iD & \sqrt{I+D^2}
\end{array}\right)
\left( \begin{array}{c|c} O^T_1 & 0 \\
\hline
 0 & O^T_2
\end{array}\right).
\end{equation}
By combining Eqs. (\ref{Y2}) and (\ref{Y1-3}) we get
\begin{equation}\label{Y}
Y=\left( \begin{array}{c|c} O_1 & 0 \\
\hline
 0 & O_2
\end{array}\right)
\left( \begin{array}{c|c} \sqrt{I+D^2} & iD \\
\hline
 -iD & \sqrt{I+D^2}
\end{array}\right)
\left( \begin{array}{c|c} O^{\prime}_1 & 0 \\
\hline
 0 & O^{\prime}_2
\end{array}\right).
\end{equation}
Finally using parameterization given in Eq. (\ref{Y2}) we get
$$
Y=\left(\begin{array}{c|c}
\begin{array}{cc}
\cosh{\theta_1} & i\sinh{\theta_1} \\
-i\sinh{\theta_1} & \cosh{\theta_1} \end{array} & 0 \\
\hline 0 & \begin{array}{cc}
\cosh{\theta_2} & i\sinh{\theta_2} \\
-i\sinh{\theta_2} & \cosh{\theta_2} \end{array}
\end{array}\right)
\left( \begin{array}{c|c} \begin{array}{cc}
\cosh{\xi_1} & 0 \\
0 & \cosh{\xi_2} \end{array} &
\begin{array}{cc}
i\sinh{\xi_1} & 0 \\
0 & i\sinh{\xi_2} \end{array} \\
\hline
 \begin{array}{cc}
-i\sinh{\xi_1} & 0 \\
0 & -i\sin{\xi_2} \end{array} &
\begin{array}{cc}
\cosh{\xi_1} & 0 \\
0 & \cosh{\xi_2} \end{array}
\end{array}\right)
$$
\begin{equation}
\left(\begin{array}{c|c}
\begin{array}{cc}
\cosh{\phi_1} & i\sinh{\phi_1} \\
-i\sinh{\phi_1} & \cosh{\phi_1} \end{array} & 0 \\
\hline 0 & \begin{array}{cc}
\cosh{\phi_2} & i\sinh{\phi_2} \\
-i\sinh{\phi_2} & \cosh{\phi_2} \end{array}
\end{array}\right),
\end{equation}
where $\sinh{\xi_i}$ (for $i=1,2$) are diagonal elements of $D$
with the conditions $\xi_i\geq0$.

Using above results and Eq. (\ref{X}) and (\ref{YX}) we can
evaluate the states $\left|x_i\right>$ as

\begin{equation}\label{x1}
{\small\left|x_1\right>=\sqrt{\frac{\lambda_1}{2}} \left(
\begin{array}{c}
-(\sinh{\xi_1}\sinh{\theta_2}\cosh{\phi_1}+\sinh{\xi_2}\cosh{\theta_2}\sinh{\phi_1})
-i(\cosh{\xi_1}\cosh{\theta_1}\cosh{\phi_1}+\cosh{\xi_2}\sinh{\theta_1}\sinh{\phi_1}) \\
-(\sinh{\xi_1}\cosh{\theta_2}\cosh{\phi_1}+\sinh{\xi_2}\sinh{\theta_2}\sinh{\phi_1})
-i(\cosh{\xi_1}\sinh{\theta_1}\cosh{\phi_1}+\cosh{\xi_2}\cosh{\theta_1}\sinh{\phi_1})
\\
(\sinh{\xi_1}\cosh{\theta_2}\cosh{\phi_1}+\sinh{\xi_2}\sinh{\theta_2}\sinh{\phi_1})
-i(\cosh{\xi_1}\sinh{\theta_1}\cosh{\phi_1}+\cosh{\xi_2}\cosh{\theta_1}\sinh{\phi_1})
\\
(\sinh{\xi_1}\sinh{\theta_2}\cosh{\phi_1}+\sinh{\xi_2}\cosh{\theta_2}\sinh{\phi_1})
-i(\cosh{\xi_1}\cosh{\theta_1}\cosh{\phi_1}+\cosh{\xi_2}\sinh{\theta_1}\sinh{\phi_1})
\end{array}
\right)},
\end{equation}
\begin{equation}\label{x2}
{\small\left|x_2\right>=\sqrt{\frac{\lambda_2}{2}} \left(
\begin{array}{c}
(\cosh{\xi_1}\cosh{\theta_1}\sinh{\phi_1}+\cosh{\xi_2}\sinh{\theta_1}\cosh{\phi_1})
-i(\sinh{\xi_1}\sinh{\theta_2}\sinh{\phi_1}+\sinh{\xi_2}\cosh{\theta_2}\cosh{\phi_1})
\\
(\cosh{\xi_1}\sinh{\theta_1}\sinh{\phi_1}+\cosh{\xi_2}\cosh{\theta_1}\cosh{\phi_1})
-i(\sinh{\xi_1}\cosh{\theta_2}\sinh{\phi_1}+\sinh{\xi_2}\sinh{\theta_2}\cosh{\phi_1})
\\
(\cosh{\xi_1}\sinh{\theta_1}\sinh{\phi_1}+\cosh{\xi_2}\cosh{\theta_1}\cosh{\phi_1})
+i(\sinh{\xi_1}\cosh{\theta_2}\sinh{\phi_1}+\sinh{\xi_2}\sinh{\theta_2}\cosh{\phi_1})
\\
(\cosh{\xi_1}\cosh{\theta_1}\sinh{\phi_1}+\cosh{\xi_2}\sinh{\theta_1}\cosh{\phi_1})
+i(\sinh{\xi_1}\sinh{\theta_2}\sinh{\phi_1}+\sinh{\xi_2}\cosh{\theta_2}\cosh{\phi_1})
\end{array}
\right)},
\end{equation}
\begin{equation}\label{x3}
{\small\left|x_3\right>=\sqrt{\frac{\lambda_3}{2}} \left(
\begin{array}{c}
(\sinh{\xi_1}\cosh{\theta_1}\cosh{\phi_2}+\sinh{\xi_2}\sinh{\theta_1}\sinh{\phi_2})
-i(\cosh{\xi_1}\sinh{\theta_2}\cosh{\phi_2}+\cosh{\xi_2}\cosh{\theta_2}\sinh{\phi_2})
\\
(\sinh{\xi_1}\sinh{\theta_1}\cosh{\phi_2}+\sinh{\xi_2}\cosh{\theta_1}\sinh{\phi_2})
-i(\cosh{\xi_1}\cosh{\theta_2}\cosh{\phi_2}+\cosh{\xi_2}\sinh{\theta_2}\sinh{\phi_2})
\\
(\sinh{\xi_1}\sinh{\theta_1}\cosh{\phi_2}+\sinh{\xi_2}\cosh{\theta_1}\sinh{\phi_2})
+i(\cosh{\xi_1}\cosh{\theta_2}\cosh{\phi_2}+\cosh{\xi_2}\sinh{\theta_2}\sinh{\phi_2})
\\
(\sinh{\xi_1}\cosh{\theta_1}\cosh{\phi_2}+\sinh{\xi_2}\sinh{\theta_1}\sinh{\phi_2})
+i(\cosh{\xi_1}\sinh{\theta_2}\cosh{\phi_2}+\cosh{\xi_2}\cosh{\theta_2}\sinh{\phi_2})
\end{array}
\right)},
\end{equation}
\begin{equation}\label{x4}
{\small\left|x_4\right>=\sqrt{\frac{\lambda_4}{2}} \left(
\begin{array}{c}
(\cosh{\xi_1}\sinh{\theta_2}\sinh{\phi_2}+\cosh{\xi_2}\cosh{\theta_2}\cosh{\phi_2})
+i(\sinh{\xi_1}\cosh{\theta_1}\sinh{\phi_2}+\sinh{\xi_2}\sinh{\theta_1}\cosh{\phi_2})
\\
(\cosh{\xi_1}\cosh{\theta_2}\sinh{\phi_2}+\cosh{\xi_2}\sinh{\theta_2}\cosh{\phi_2})
+i(\sinh{\xi_1}\sinh{\theta_1}\sinh{\phi_2}+\sinh{\xi_2}\cosh{\theta_1}\cosh{\phi_2})
\\
-(\cosh{\xi_1}\cosh{\theta_2}\sinh{\phi_2}+\cosh{\xi_2}\sinh{\theta_2}\cosh{\phi_2})
+i(\sinh{\xi_1}\sinh{\theta_1}\sinh{\phi_2}+\sinh{\xi_2}\cosh{\theta_1}\cosh{\phi_2})
\\
-(\cosh{\xi_1}\sinh{\theta_2}\sinh{\phi_2}+\cosh{\xi_2}\cosh{\theta_2}\cosh{\phi_2})
+i(\sinh{\xi_1}\cosh{\theta_1}\sinh{\phi_2}+\sinh{\xi_2}\sinh{\theta_1}\cosh{\phi_2})
\end{array}
\right)}.
\end{equation}
Equations (\ref{x1}) to (\ref{x4}) together with normalization
condition $\sum_{i=1}^4\left<x_i|x_i\right>=1$ give a
parameterization for a generic orbit of two qubit density matrix
up to local unitary group. As an example let us consider Bell
decomposable states
$\rho=\sum_{i=1}^{4}p_{i}\left|\psi_i\right>\left<\psi_i\right|$,
where $p_i\geq0,\,\sum_i p_i=1$. For these states by choosing
$\theta_1=\theta_2=\xi_1=\xi_2=\phi_1=\phi_2=0$ we get
$\lambda_i=p_i$ and states $\left|x_i\right>$ are given by
\begin{eqnarray}
\left|x_1\right>=-i\sqrt{p_1}\left|\psi_1\right>,\qquad
\left|\psi_1\right>
=\frac{1}{\sqrt{2}}(\left|\uparrow\uparrow\right>+\left|
\downarrow\downarrow\right>),
\\
\left|x_2\right>=\sqrt{p_2}\left|\psi_2\right>,\qquad
\left|\psi_2\right>
=\frac{1}{\sqrt{2}}(\left|\uparrow\downarrow\right>+\left|
\downarrow\uparrow\right>),
\\
\left|x_3\right>=-i\sqrt{p_3}\left|\psi_3\right>,\qquad
\left|\psi_3\right>
=\frac{1}{\sqrt{2}}(\left|\uparrow\downarrow\right>-\left|
\downarrow\uparrow\right>),
\\
\left|x_4\right>=\sqrt{p_4}\left|\psi_4\right>,\qquad
\left|\psi_4\right>
=\frac{1}{\sqrt{2}}(\left|\uparrow\uparrow\right>-\left|
\downarrow\downarrow\right>).
\end{eqnarray}

\section{Conclusion}
We have obtained  Lewenstein-Sanpera decomposition for a generic
two qubit density matrix by using Wootters's basis. It is shown
that the average concurrence of the decomposition is equal to the
concurrence of the state. It is also shown that all entanglement
content of the state is concentrated in the Wootters's state
$\left|x_1\right>$ associated with the largest eigenvalue
$\lambda_1$. It is shown that a given density matrix $\rho$ with
corresponding set of positive numbers $\lambda_i$ and Wootters's
basis can transforms under $SO(4,c)$ into a generic $2\times2$
matrix with the same set of positive numbers but with new
Wootters's basis. We have also shown that the local unitary
transformations correspond to $SO(4,r)$ transformations, hence,
$\rho$ can be represented as coset space $SO(4,c)/SO(4,r)$
together with positive numbers $\lambda_i$.

{\large \bf Appendix }

Let us consider the set of linearly independent vectors
$\{\left|\phi_i\right>\}$, then one can define their dual vectors
$\{\left|{\hat \phi}_i\right>\}$ such that the following relation
\begin{equation}\label{A1}
\left<{\hat \phi}_i\mid\phi_j\right>=\delta_{ij}
\end{equation}
hold. It is straightforward to show that the
$\{\left|\phi_i\right>\}$ and their dual $\{\left|{\hat
\phi}_i\right>\}$ posses the following completeness relation
\begin{equation}\label{ids}
\sum_{i}|{\hat \phi}_i\left>\right<\phi_i|=I,
\qquad\sum_{i}|\phi_i\left>\right<{\hat \phi}_i|=I.
\end{equation}
Consider an invertible     operator $M$ which is expanded in terms
of states $\left|\phi_i\right>$ as
\begin{equation}\label{M}
M=\sum_{i}a_{ij}\left|\phi_i\right>\left<\phi_j\right|
\end{equation}
Then the inverse of $M$ denoted by $M^{-1}$ can be expanded in
terms of dual bases as
\begin{equation}\label{Minverse}
M^{-1}=\sum_{i}b_{ij}|{\hat \phi}_i\left>\right<{\hat \phi}_j|
\end{equation}
where $b_{ij}=(A^{-1})_{ij}$ and $A_{ij}=a_{ij}$.

\end{document}